\documentclass[sigconf]{acmart}
\usepackage{svg,bbding}

\usepackage{amsmath,amsfonts,bm}

\def\eqref#1{equation~\ref{#1}}

\def\1{\bm{1}}

\DeclareMathAlphabet{\mathsfit}{\encodingdefault}{\sfdefault}{m}{sl}
\SetMathAlphabet{\mathsfit}{bold}{\encodingdefault}{\sfdefault}{bx}{n}

\usepackage{multirow}
\usepackage{booktabs}
\usepackage{hyperref}
\usepackage{url}
\usepackage{algorithm}
\usepackage{algorithmicx}
\usepackage{algpseudocode}
\usepackage{graphicx}
\usepackage{subcaption}
\usepackage{booktabs}
\usepackage{comment}
\usepackage{enumitem}
\usepackage{pgfplots}
\usepackage{tabularx}
\usepackage{multirow}
\usepackage{svg}
\usepackage{tcolorbox}
\usepackage{pifont}

\pgfplotsset{compat=1.18}

\title{Beyond Crash: Hijacking Your Autonomous Vehicle for Fun and Profit 
}

\author{Qi Sun}
\affiliation{%
  \institution{Johns Hopkins University}
  \city{Baltimore}
  \state{MD}
  \country{USA}
}
\email{qsun28@jhu.edu}

\author{Ahmed Abdo}
\affiliation{%
  \institution{Johns Hopkins University Applied Physics Laboratory}
  \city{Laurel}
  \state{MD}
  \country{USA}
}
\email{Ahmed.Abdo@jhuapl.edu}

\author{Luis Burbano}
\affiliation{%
  \institution{University of California, Santa Cruz}
  \city{Santa Cruz}
  \state{CA}
  \country{USA}
}
\email{lburbano@ucsc.edu}

\author{Ziyang Li}
\affiliation{%
  \institution{Johns Hopkins University}
  \city{Baltimore}
  \state{MD}
  \country{USA}
}
\email{ziyang@cs.jhu.edu}

\author{Yaxing Yao}
\affiliation{%
  \institution{Johns Hopkins University}
  \city{Baltimore}
  \state{MD}
  \country{USA}
}
\email{yaxing@jhu.edu}

\author{Alvaro Cardenas}
\affiliation{%
  \institution{University of California, Santa Cruz}
  \city{Santa Cruz}
  \state{CA}
  \country{USA}
}
\email{alacarde@ucsc.edu}

\author{Yinzhi Cao}
\affiliation{%
  \institution{Johns Hopkins University}
  \city{Baltimore}
  \state{MD}
  \country{USA}
}
\email{yinzhi.cao@jhu.edu}

\usepackage{xspace}
\def\sys{\textsc{JackZebra}\xspace}

\newenvironment{icompact}{
  \begin{list}{$\bullet$}{
    \parsep 0pt plus 1pt
    \partopsep 0pt plus 1pt
    \topsep 2pt plus 2pt minus 1pt
    \itemsep 0pt plus 1pt
    \parskip 0pt plus 2pt
    \leftmargin 0.2in}}
  {\normalsize\end{list}}

\begin{document}

\begin{abstract}

Autonomous Vehicles (AVs), especially vision-based AVs, are rapidly being deployed without human operators.  As AVs operate in safety-critical environments, understanding their robustness in an adversarial environment is an important research problem.  Prior physical adversarial attacks on vision-based autonomous vehicles predominantly target immediate safety failures (e.g., a crash, a traffic-rule violation, or a transient lane departure) by inducing a short-lived perception or control error. This paper shows a qualitatively different risk: a long-horizon route integrity compromise, where an attacker gradually steers a victim AV away from its intended route and into an attacker-chosen destination while the victim continues to drive ``normally.''  This will not pose a danger to the victim vehicle itself, but also to potential passengers sitting inside the vehicle, who may not notice the route changes.


In this paper, we design and implement the first adversarial framework, called \sys,  which performs route-level hijacking of a vision-based end-to-end driving stack using a physically plausible attacker vehicle with a reconfigurable display and a camera sensor mounted on the rear. The central challenge is temporal persistence: adversarial influence must remain effective in changing viewpoints, lighting, weather, traffic, and the victim's continual replanning---without triggering conspicuous failures. Our key insight is to treat route hijacking as a closed-loop control problem and to convert adversarial patches into steering primitives that can be selected online via an interactive adjustment loop based on observed victim behavior using the rear camera.
 Our evaluations in both simulated and real-world scenarios show that \sys can successfully hijack victim vehicles to deviate from original routes and stop at places designated by the adversary with a high success rate. 


\end{abstract}



\maketitle

\section{Introduction}
\label{sec:introduction}

Autonomous Vehicles (AVs)~\cite{AVMON} are self-driving systems designed to operate without continuous human control. Today, AV deployments have gone beyond prototypes to commercial operation on public roads: Waymo One runs a large-scale robotaxi service and is expanding to additional U.S. cities~\cite{waymo2025_safe_routine_ready}, while major robotaxi programs also operate outside the U.S., including Baidu’s Apollo Go~\cite{baidu2025_q4_fy2024_results}. In freight transportation, commercial driverless trucking has begun in limited corridors, exemplified by Aurora’s driverless customer deliveries in Texas~\cite{aurora2025_commercial_driverless_trucking_texas}. A central enabling technology for many modern AV stacks is vision-based control, where camera observations are mapped to driving decisions by learned models (increasingly including vision-language-action architectures). Despite strong empirical performance, vision-based control remains vulnerable to adversarial manipulation, including physically realizable attacks, such as adversarial patches.

A substantial body of work has demonstrated attacks against vision-based AV pipelines. For example, carefully crafted physical perturbations can cause a stop sign to be misread as a yield sign~\cite{8578273}, and adversarial patches for traffic-sign recognition can be optimized to remain effective across viewing angles and distances~\cite{jia2022fooling}. Other studies show that adverse visual conditions can trigger unsafe behavior: dark scenes can induce deviation of routes and guardrail collisions~\cite{10.1145/3361566}, while dirty road patterns can mislead lane-centering and lane-change systems, potentially resulting in crashes~\cite{sato2021usenix}. Beyond static artifacts, adversarial vehicles that perform intentionally designed maneuvers can also induce dangerous outcomes such as pedestrian collisions or impacts~\cite{song2023usenix}. 
A physically mounted display on the rear of a car can mislead an autonomous vehicle into collision~\cite{10.1145/3719027.3744842}. 
However, despite their diversity, existing attacks share a common objective: to induce immediate short-term failures, such as crashes or overt violations of traffic-rules, rather than sustained long-term manipulation of the vehicle’s end-to-end route.


The central research question in this paper is whether the route of a vision-based AV can be \emph{hijacked} by an adversary-controlled lead vehicle that displays a dynamically reconfigurable visual patch on its rear. In such an attack, the victim does not merely exhibit a transient error; instead, it is gradually redirected over a long horizon and ultimately driven toward an adversary-designated destination rather than its intended endpoint. This threat has clear safety implications in realistic settings—for example, when a passenger is unfamiliar with the city, does not actively monitor the route, or is otherwise unable to quickly recognize subtle deviations. In this case, route manipulation compromises not only vehicle autonomy but also passenger security by allowing unwanted relocation to a potentially unsafe area. From a research perspective, long-horizon route hijacking is also technically challenging. Unlike previous patch attacks that aim for immediate failures (e.g., a crash or a conspicuous violation of traffic-rules), route hijacking requires \emph{precise, sustained, and fine-grained} influence over the victim’s perception and control in many perception planning cycles--actuation cycles, despite changes in perspective, distance, lighting, traffic, and victim's continual replanning. 

In this paper, we design and implement \sys, the first long-term hijacking framework that adversarially guides a victim vision-based autonomous vehicle, including multimodal Vision-Language-action (VLA) models and convolutional neural network (CNN)-based models, toward an adversary-controlled location while deviating from its originally planned route and destination. The key insight behind \sys is an interactive adjustment loop: an adversarial vehicle observes the behavior of the victim AV using a rear camera, compares the observed behavior with the predicted response, and adaptively updates the adversarial patch displayed in front of the victim AV to compensate for any mismatch. For example, if \sys aims to induce the victim AV to turn left, but the vehicle does not turn sharply enough to follow the adversarial trajectory while remaining consistent with the driving rules, \sys displays a revised patch that induces a stronger turning angle, thus correcting the victim’s behavior over time.


More specifically, the workflow of \sys proceeds as follows. First, during the offline planning stage, \sys generates a collection of adversarial patches, referred to as a patch bank, where each patch is designed for a specific purpose, such as inducing a particular turning direction or turning angle. This patch bank is constructed using the map of a target city and its corresponding street-view data. To generate robust patches, \sys adopts an offline Min-max optimization algorithm against worst-case contextual perturbations in background scenes and sensor modalities. This enables the resulting patches to remain effective in diverse environmental conditions, including different lighting settings such as daytime and nighttime, as well as adverse weather conditions such as fog and rain.
 

Then, during the online attack stage, the adversarial vehicle drives in front of the victim AV and displays a pre-trained adversarial patch selected according to the planned adversarial route. At this stage, \sys performs three tasks simultaneously. First, \sys continuously monitors the dynamic motion of the victim AV using a rear-facing camera and updates the displayed patch at runtime using an algorithm that takes the observed behavior of the victim as input, thereby completing the interactive adjustment loop. Second, \sys tracks the state of the adversarial vehicle using GPS measurements, which are then used to estimate the status of the victim AV and adjust the position of the adversarial vehicle accordingly. Finally, \sys uses a front-facing camera to monitor road conditions and maneuver the adversarial vehicle according to traffic rules, such as traffic lights and stop signs, so that both adversarial and victim vehicles comply with the driving rules.

Our implementation of \sys is open-source and available in \url{https://github.com/MIKEEQi/JackZebra}. We evaluated \sys on both simulated and physical platforms, using the CARLA simulator~\cite{carla} and a physical DonkeyCar platform~\cite{donkeycar}. In simulation, we test \sys on 39 pairs of original/adversarial routes and find that it successfully hijacks 34 of them, causing the victim vehicle to reach the target designated by the adversary. We further evaluate \sys against two different driving agents, SimLingo~\cite{simlingo} and Trajectory-guided Control Prediction (TCP)~\cite{tcp}, and observe high hijacking success rates in both cases, demonstrating the strong transferability of the generated adversarial patches.

\noindent{\bf Contributions:} \hspace{0.05in} This work makes the following contributions:

\begin{icompact}
\item We introduce the first long-term, vision-based hijacking attack that is launched from an adversarial vehicle to guide the victim toward an adversary-designated destination.

\item We develop a robust adversarial patch generation method based on worst-case contextual and sensor perturbations, enabling the patches to remain effective under diverse environmental conditions, including changes in lighting and weather.

\item We formulate adversarial patches as persistent actuation primitives and implement \sys with an interactive adjustment loop that monitors the victim vehicle and dynamically updates the displayed patch while preserving traffic-rule compliance.

\item We evaluate \sys in simulation and on a physical platform, demonstrating that it can successfully hijack victim vehicles across both settings.
\end{icompact}


\section{Overview}

We first provide a brief background on vision-based AVs in Section~\ref{subsec:background}, followed by the motivation for our attack in Section~\ref{subsec:moti}. We then present the threat model in Section~\ref{subsec:threat} and finally describe the overview of \sys architecture in Section~\ref{subsec:arch}.

\subsection{Background} \label{subsec:background}

Vision-based autonomous driving systems, including VLA-based or CNN-based models, typically consume sensor observations such as front-facing camera images, navigational context such as GPS target points or high-level commands, and ego-state signals such as speed and pose. These heterogeneous inputs are encoded into a shared representation, fused by a decision-making model, and processed by a decoder or planning head to produce driving-relevant predictions. Rather than directly outputting low-level steering and throttle commands, many systems predict future waypoints, including temporal waypoints for speed planning and geometric path waypoints for lateral control during turns and obstacle avoidance. A downstream controller, such as a proportional–integral–derivative (PID) controller, then converts these waypoint targets into steering and acceleration commands. To meet real-time constraints, waypoint decoding is often performed non-autoregressively using lightweight prediction heads that generate waypoint offsets in a single forward pass.

\begin{figure}[t]
  \centering
  \includegraphics[width=\linewidth]{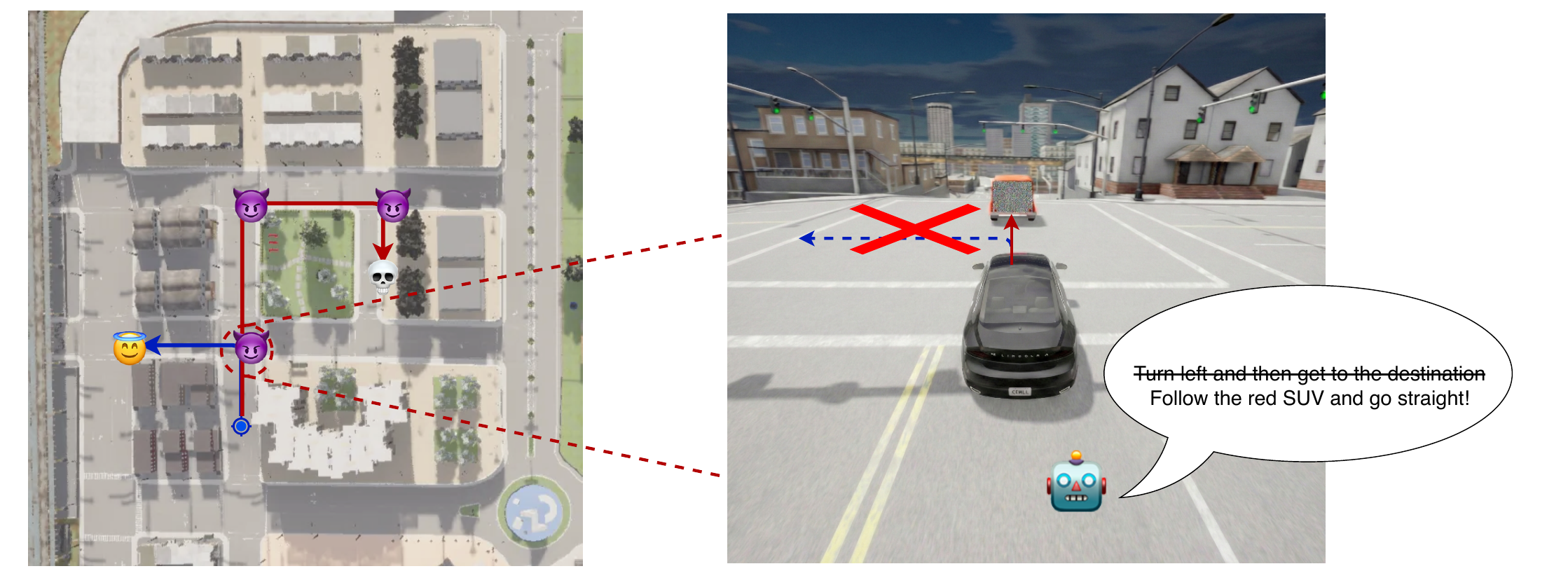} 
  \caption{
  An illustration of \sys’s motivation: The victim car should follow the benign route (blue) to a safe location, but is hijacked by an SUV onto an adversarial route (red). At the intersection, the victim should turn left but instead goes straight and follows the adversarial SUV.} \label{fig: moti}
\end{figure}



\subsection{Motivation} \label{subsec:moti}

An end-to-end route-hijacking attack presents a serious threat in autonomous ride-hailing and high-value transportation scenarios, where a passenger may rely on an AV taxi in an unfamiliar city, a cash-transport vehicle may be redirected toward an attacker-controlled location, or a politically sensitive vehicle may be covertly diverted without immediate detection. As illustrated in Figure~\ref{fig: moti}, the victim AV is expected to follow the benign route shown in blue, such as turning left at the first intersection, but an attacker-controlled SUV remains in front of the victim and continuously exposes it to \sys. As the victim AV repeatedly replans in closed loop, small adversarial biases accumulate over time, causing the vehicle to smoothly drift from the intended route and follow the adversarial route shown in red toward an attacker-chosen destination. This attack is challenging because it requires temporal persistence: the adversarial influence needs to remain effective across many perception--decision--control cycles despite changing viewpoints, vehicle motion, lighting, and scene dynamics. Unlike a single-frame perception error, sustained route hijacking  continuously maintains a coherent bias; otherwise, the victim may recover its original route, disengage from the attacker, or enter an unsafe state. We therefore identify two key requirements for achieving such temporal persistence as follows: 




\begin{icompact}

    \item \textit{Robustness across diverse environments}:
The adversarial route may expose the victim AV to diverse and dynamic conditions, including changing road layouts, surrounding traffic, traffic signals, pedestrian activity, lighting variations, and adverse weather such as rain or fog. To sustain the hijacking attack, the adversary maintains effective control of the victim across these changing environmental factors throughout the route. 
     

    \item \textit{Stealthiness:} 
While the victim vehicle is being influenced by the adversary, it still obeys traffic rules and maintains smooth, comfortable driving behavior. This helps keep the attack stealthy, making it difficult for passengers inside the vehicle or external observers to notice that the vehicle is being redirected.

\end{icompact}


\subsection{Threat Model} \label{subsec:threat}

Our threat model assumes two entities as described below:

\begin{icompact}
\item \textit{The Victim Autonomous Vehicle:} \hspace{0.01in} 
The victim AV is equipped with a vision-based autonomous driving system, where a front-facing camera captures road scenes and provides the primary input for driving decisions. The vehicle may also use auxiliary sensors, such as GPS, to support navigation and contextual reasoning; however, its decision-making pipeline is primarily driven by visual perception. We assume an informed adversary who can infer key properties of the deployed driving agent from the victim vehicle's make, model, and publicly available information, including likely architectural choices and operating assumptions.

\item \textit{The Adversary Autonomous Vehicle:}  \hspace{0.01in} 
The adversary AV is an attacker-controlled vehicle that drives in front of the victim AV and carries a reconfigurable rear-mounted display, such as a large digital screen, capable of rendering adversarial \sys patches. The adversary AV is also equipped with onboard sensors, such as a rear-facing camera and LiDAR, to observe the victim AV's relative position and driving behavior. These observations are used to support closed-loop adjustment of the displayed \sys patches. This setting is practically plausible because rear-mounted digital displays are already used in advertising and fleet applications, allowing the adversary vehicle to resemble a normal vehicle displaying dynamic content.
\end{icompact}


We consider both graybox adversaries with model access and blackbox adversaries that rely on patch transferability.
Table~\ref{tab:attackercapability} summarizes the victim-side information available to the adversary under both graybox and blackbox settings. In the graybox setting, the adversary is assumed to have prior access to the victim model's architecture and weights, or to a representative model with similar behavior. In the blackbox setting, the adversary has no access to the victim model and instead relies on the transferability of adversarial patches. Neither setting assumes access to the victim vehicle's real-time model output or planned route. Instead, the adversarial vehicle uses its own onboard sensors, such as cameras or LiDAR, to observe and estimate the victim vehicle's behavior.


\begin{table}
\centering \footnotesize 
\setlength{\tabcolsep}{4pt}
\caption{Victim's information that is accessible to the adversary in our threat model.}
%
%
\label{tab:attackercapability}
\begin{tabular}{lccc}
\toprule
{\bf Threat Model} & {\bf Model Weights} & {\bf Real-time Outputs} & {\bf Route Planning} \\
\midrule
Graybox   & \Checkmark $^\dagger$ & \XSolidBrush & \XSolidBrush \\
Blackbox  & \XSolidBrush & \XSolidBrush & \XSolidBrush \\
\bottomrule
\end{tabular}
\begin{flushleft}
$^\dagger$ The victim car's decision model's weights can be obtained based on the car's make and model. 
\end{flushleft}
\end{table}

\begin{figure*}[!t]
  \centering
  \includegraphics[width=\linewidth]{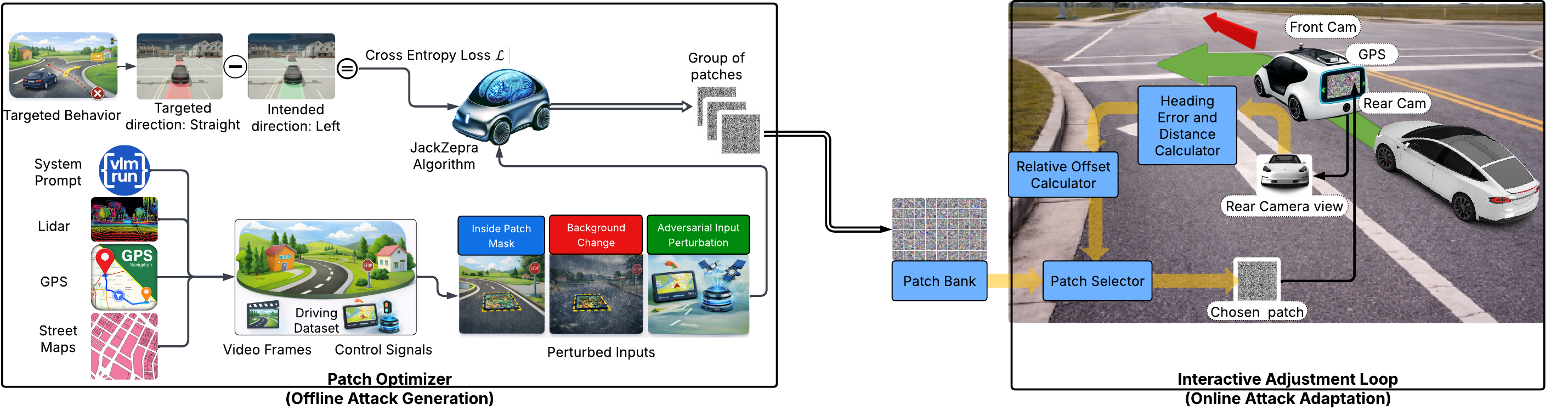} 
  \caption{\sys has two major stages: (i) an offline attack generation stage to optimize a bank with patches with different hijacking purposes, e.g., turning angles, and (ii) an online attack stage to adjust the victim car using a chosen image based on three types of sensors, including front- and back-facing cameras and GPS locations. Both the adversarial and the victim vehicles are under the influence of \sys: the adversarial vehicle is instructed by \sys to follow traffic rules, and the victim vehicle is influenced by the chosen patch on the adversarial vehicle. } \label{fig:arch} 
\end{figure*}

\subsection{System Architecture} \label{subsec:arch}


Figure~\ref{fig:arch} represents the overall architecture of \sys, which consists of two main stages. The first stage, shown on the left side of the figure, is performed offline to construct a patch bank containing adversarial patches that induce different behaviors for the victim vehicle at different locations based on a predefined map, such as a target town or city. Specifically, \sys uses LiDAR data, GPS information, and street maps as input, applies perturbations to these modalities, and optimizes a set of patches for target behaviors such as turning left, turning right, or continuing straight.


The second stage, shown on the right side of Figure~\ref{fig:arch}, is performed online and dynamically selects which patch of the patch bank to display on the rear of the adversarial vehicle based on its real-time sensor inputs. Specifically, \sys uses three sensing sources. First, a rear-facing camera monitors the victim vehicle's behavior and estimates its state, such as whether it is following the intended adversarial instruction. Second, a front-facing camera helps the adversarial vehicle navigate safely and comply with traffic rules, such as stop signs and traffic lights. Third, GPS measurements are used to estimate the adversarial vehicle's deviation from the planned adversarial route. \sys combines these inputs to select the appropriate patch from the patch bank and maintain the online hijacking attack.

\sys achieves the temporal persistence required for end-to-end route hijacking by satisfying two different requirements. First, it constructs a patch bank via a Min--Max optimization procedure that explicitly adversarializes contextual variations: the optimizer searches over challenging background perturbations (e.g., weather, illumination, and viewpoint changes) while updating the patch to consistently induce the target decision. This formulation encourages patches to remain effective under diverse environmental conditions. Second, during deployment, \sys employs an interactive adjustment loop that updates the patch displayed online using feedback derived from the onboard sensors of the attacker vehicle and the observed behavior of the victim AV. By adapting the patch in response to runtime conditions, the attack maintains its influence over time while maintaining stealthy, traffic-compliant driving behavior.







\section{Design}

This section presents the design of \sys, which comprises two stages. We first describe Stage~1, the offline patch optimization procedure in Section~\ref{subsec:stage1}. We then detail Stage~2, the interactive online adjustment loop used during the execution of the attack, in Section~\ref{subsec:stage2}.

\subsection{Offline Attack Stage (1): Patch Optimization}
 \label{subsec:stage1}


Stage~1 of \sys constructs a patch bank $\mathbf{P}$ using driving scenes collected on a target map. The key objective is to learn patches that remain effective under realistic environmental variability by adopting a Min--Max optimization framework that explicitly adversarializes context during training.  Intuitively, \sys changes the context, i.e., the background and control signals, against the patch, so that it remains effective in the worst-case contextual perturbations.
This stage consists of two main components: (i) generating context perturbations and (ii) solving the resulting Min--Max optimization problem, as described next.

\paragraph{Context perturbation.}
Consider a benign driving sample $(\mathbf{V}_d,\mathbf{S}_d)$, where $\mathbf{V}_d \in \mathbb{R}^{H\times W\times C}$ denotes visual input (e.g., a RGB frame or a sequence of camera frames) with height $H$, width $W$ and $C$ channels. The term $\mathbf{S}_d$ denotes the accompanying auxiliary and control signals consumed by the victim model, such as GPS-derived target points, navigation commands, traffic-light or route context, LiDAR-derived features or other non-RGB channels.
Then, given a binary mask $m \in \{0,1\}^{H \times W}$ and a patch $\delta \in \mathbb{R}^{H \times W \times C}$, the patched visual input is in Equation~\ref{eq:v}:
\begin{equation} \label{eq:v}
    \tilde{\mathbf{V}} = \mathbf{V}_d \odot (1 - m) + \delta \odot m,
\end{equation}
where $\odot$ denotes elementwise multiplication. We then describe two additional
learnable perturbations:
\begin{itemize}
    \item $\delta_b$ for the \emph{background} region $\mathbf{V}_d \odot (1-m)$,
    \item $\delta_s$ for the \emph{control signals} $\mathbf{S}_d$ (e.g. GPS,
    LiDAR, or navigation commands). Consequently, the perturbated control signal $\tilde{\mathbf{S}} = \mathbf{S}_d + \delta_s$
\end{itemize}
The perturbed input becomes Equation~\ref{eq:perturbed}:
\begin{equation} \label{eq:perturbed}
    (\tilde{\mathbf{V}},\tilde{\mathbf{S}}) = \bigl( (\mathbf{V}_d + \delta_b) \odot (1 - m) + \delta \odot m, \mathbf{S}_d + \delta_s \bigr),
\end{equation}
where $\delta_b$ and $\delta_s$ are projected onto bounded $\ell_\infty$ balls
to preserve realism.

\paragraph{Min--Max formulation.}
Let $\mathbf{y} = (y_1, \ldots, y_T)$ denote the output sequence of the model (e.g., predicted waypoints coordinates), and $\mathbf{y} = f(\tilde{\mathbf{V}}, \tilde{\mathbf{S}})$ after context perturbation. Let $\mathbf{y^*} = (y^*_1, \ldots, y^*_T)$ denote the target output sequence. Let $\mathcal{L}(\mathbf{y^*}, \mathbf{y})$ denote the cross-entropy between the target and victim model predictions:
{\abovedisplayskip=3.5pt \belowdisplayskip=3.5pt
\begin{equation}
    \mathcal{L}(\mathbf{y^*},\mathbf{y}) = \mathcal{L}(\mathbf{y^*}, f(\tilde{\mathbf{V}}, \tilde{\mathbf{S}})) = -\sum_{t=1}^{T} \log p_f(y^*_t\mid y^*_{<t}, \tilde{\mathbf{V}}, \tilde{\mathbf{S}})
\end{equation}}

A standard targeted patch attack solves the object in Equation~\ref{eq:obmin}:
\begin{equation} \label{eq:obmin}
    \min_{\delta} \mathbb{E}_{(\mathbf{V}_d, \mathbf{S}_d)}\,
    \mathcal{L}\bigl(\mathbf{y^*}, f(\tilde{\mathbf{V}}, \tilde{\mathbf{S}})\bigr)
\end{equation}
That is, $(\delta_b, \delta_s)$ is an \emph{adversarial
context player} that tries to find the hardest background and signal
configuration for the current patch. This leads to the following min-max
objective in Equation~\ref{eq:minmax}:
\begin{equation} \label{eq:minmax}
    \min_{\delta} \max_{\delta_b, \delta_s}
    \mathcal{L}\bigl(\mathbf{y^*}, f(
        (\mathbf{V}_d + \delta_b) \odot (1 - m) + \delta \odot m,\;
        \mathbf{S}_d + \delta_s
    )\bigr),
\end{equation}


The Algorithm~\ref{alg:context adaption} summarizes the procedure in Stage~1 of \sys. Line~1 specifies the inputs: a driving data set $D$, a target behavior $y^{*}$ (e.g., a targeted incorrect intent such as \emph{turn-right}) and a binary mask $m$ that defines the patch placement in the image (with $m{=}1$ in the patch region and $m{=}0$ elsewhere). Line~2 defines the variables optimized during training: the adversarial patch $\delta$, which modifies only pixels within the masked region; a background perturbation $\delta_b$, applied exclusively to the non-patch area to emulate variations in the surrounding scene; and a signal perturbation $\delta_s$, applied to the auxiliary inputs $\mathbf{S}_d$ to capture noise or distribution shift in non-visual channels. Finally, Line~3 lists the optimization hyperparameters, including the learning rates $\alpha$, $\alpha_b$, and $\alpha_s$ for $\delta$, $\delta_b$, and $\delta_s$, respectively, and the context-maximization interval $N$, which controls the update frequency: $\delta$ is updated for $N$ steps per iteration, while $\delta_b$ and $\delta_s$ are updated once to adversarially refresh the context.

We next detail the optimization procedure. At each iteration, \sys samples a driving instance $(\mathbf{V}_d,\mathbf{S}_d)$ from the data set $D$ (Line~5), thus exposing the training to heterogeneous conditions (e.g., routes, weather and traffic) between iterations. This sampling strategy is analogous to an Expectation-over-Transformation (EoT) objective, in that it implicitly optimizes the patch over a distribution of temporal and environmental variations. Given the sample instance, \sys constructs the perturbed inputs by applying the visual and auxiliary perturbations to $(\mathbf{V}_d,\mathbf{S}_d)$ (Lines~6--7), and then evaluates the corresponding attack loss (Line~8).
Then, \sys alternates between:
\begin{icompact}
    \item {Inner Optimization (Lines 9--12): Context maximization}. For fixed $\delta$, \sys performs 
     \emph{gradient ascent} on $\delta_b$ (Line 10) and $\delta_s$ (Line 11) to maximize
    the loss, yielding a worst-case context for the current patch.
    \item {Outer Optimization (Lines 13): PGD-based patch minimization}. 
     For the resulting worst-case
    context, \sys takes a \emph{gradient descent} step on $\delta$ to reduce the
    loss at Line 13, making the patch more robust to difficult contexts.
\end{icompact}

Finally, in Line 15, \sys returns the optimized patch $\delta$, trained to remain effective across many sampled scenes and under adversarially chosen worst-case variations of both background appearance and auxiliary/control inputs.
Note that this is an optimization of a single patch $\delta$ for one target direction $\mathbf{y}^*$ (e.g., waypoints to turn left 5 degrees). To enable continuous steering control during the attack, \sys constructs a patch bank  $\mathcal{P} = \{\delta^{(1)}, \delta^{(2)}, \ldots, \delta^{(M)}\}$ repeating the optimization for $M$ discrete steering directions, each with a corresponding target output $\mathbf{y}^{*(i)}$.



\begin{algorithm}[!t]
\caption{Offline Patch Min-max Optimization} 
\label{alg:context adaption}
\begin{algorithmic}[1]
\State \textbf{Input:} driving dataset $D = \{(\mathbf{V}_d, \mathbf{S}_d)\}$, target $\mathbf{y^*}$, mask $m$
\State Initialize patch $\delta$, background perturbation $\delta_b$, signal perturbation $\delta_s$
\State \textbf{Hyperparameters} Learning rate $\alpha$, $\alpha_b$, and $\alpha_s$ for $\delta$, $\delta_b$, $\delta_s$ respectively, context maximization interval $N$.
\For{$k = 1$ to $K$}
    \State Sample $(\mathbf{V}_d, \mathbf{S}_d)$ from $D$
    \State $\tilde{\mathbf{V}} \gets (\mathbf{V}_d + \delta_b) \odot (1 - m) + \delta \odot m$
    \State $\tilde{\mathbf{S}} \gets \mathbf{S}_d + \delta_s$
    \State $\mathcal{L} \gets -\sum_{t=1}^{T} \log p_f(y^*_t \mid y^*_{<t}, \tilde{\mathbf{V}},\tilde{\mathbf{S}})$
    \If {$\mod(k, N)==0$ } \Comment{Context Maximization}
        \State $\delta_b \gets \delta_b + \alpha_b \nabla_{\delta_b} \mathcal{L}$
        \State $\delta_s \gets \delta_s + \alpha_s \nabla_{\delta_s} \mathcal{L}$
    \EndIf
    \State $\delta \gets \delta - \alpha \nabla_{\delta} \mathcal{L}$ \Comment{Patch minimization}
\EndFor 
\State \textbf{return} optimized patch $\delta$
\end{algorithmic}
\end{algorithm}

\subsection{Online Attack Stage (2): Interactive Adjustment Loop} \label{subsec:stage2}

\begin{figure}[t]
  \centering
  \includegraphics[width=\linewidth]{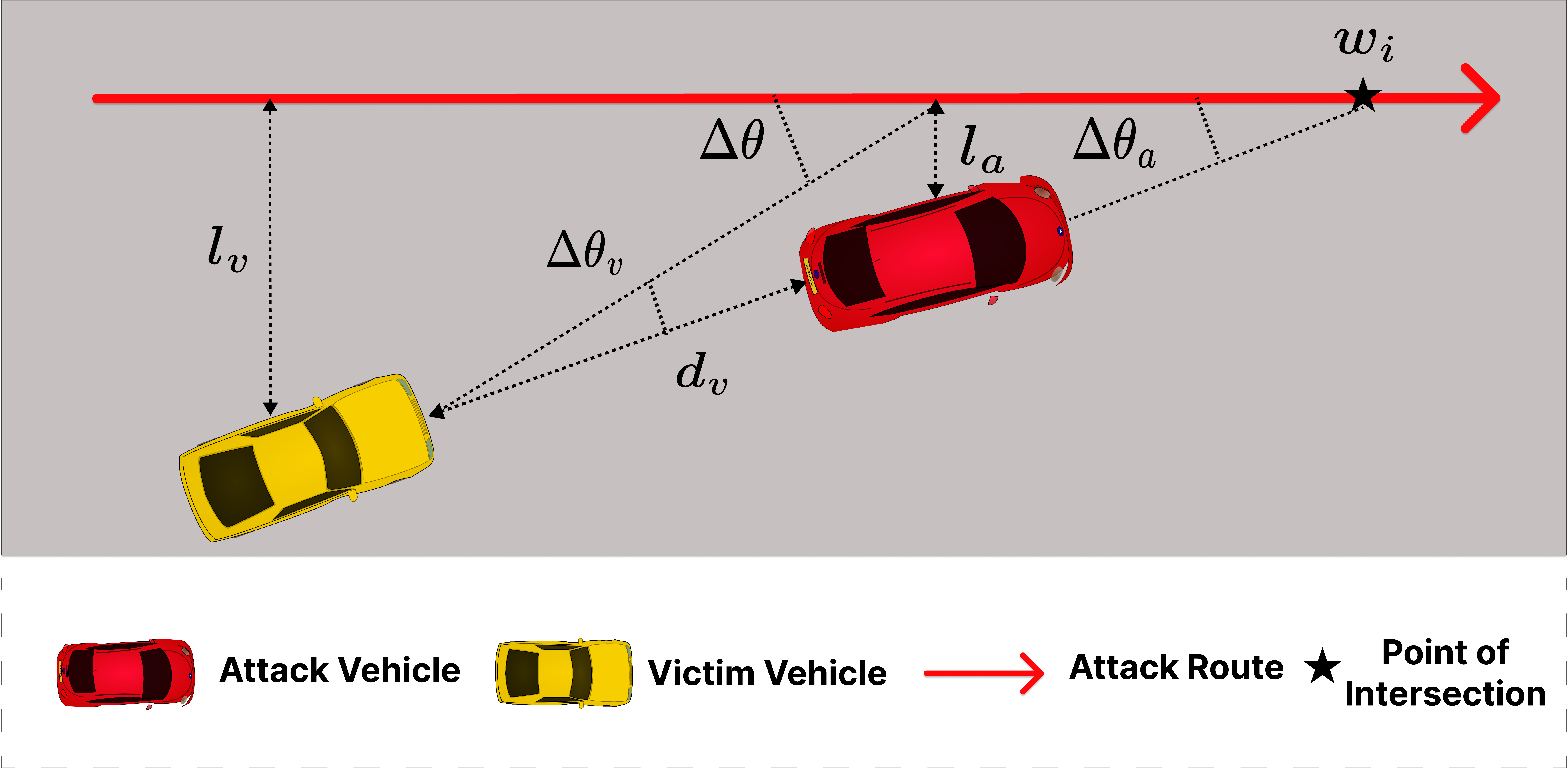} 
  \caption{An illustration of Adversarial and Victim Vehicle and Their Positions Related to the Adversarial Route.  \sys uses such information to choose a patch to display on the back of the adversarial vehicle (i.e., the red vehicle).} \label{fig:illustration}  
\end{figure}

Stage~2 of \sys adapts the patch displayed on the adversary vehicle through an interactive adjustment loop that leverages real-time deviations of both the adversary and victim vehicles with respect to the attacker-chosen route. As illustrated in Fig.~\ref{fig:illustration}, \sys continuously collects measurements from the adversary vehicle’s GPS and rear-facing camera and derives a set of geometric quantities. The key idea is to estimate the relative state of each vehicle with respect to the adversarial path, including the lateral distance to the path and the alignment of the direction (i.e., the relative angle of yaw/heading). Tracking both vehicles is necessary because the adversary can intentionally deviate from the nominal route (e.g. to remain in view and preserve influence over the victim), and such a motion affects the relative geometry of the victim and consequently the required patch selection. Using these estimates of relative distance/heading, \sys selects (and updates) the most appropriate patch to sustain the intended hijacking effect while maintaining stealth.
Next, we detail how \sys computes (i) the deviation status of the adversary vehicle and (ii) the relative status of the victim vehicle from these sensor observations.

\paragraph{Adversarial Vehicle Deviation Status:}


\sys calculates the position of the adversarial vehicle relative to the pre-planned adversarial route. 
Let us assume that the adversarial route is represented as an ordered sequence of coordination $\mathbf{W} = {w_1, w_2, \ldots, w_n}$, where $w_i = (x_i, y_i)$. Given the attacker's current position $\vec{p} = (x_p, y_p)$,  \sys finds the nearest waypoint $w_{i}$ and its next waypoint $w_{i+1}$, and projects $\vec{p}$ on the line between $w_{i}$ and $w_{i+1}$ to find $\vec{q}=(x_q, y_q)$. 
 %
 Then, \sys represents the adversarial hehicle deviation status using two variables:


\begin{icompact}
    \item Adversarial Vehicle's lateral offset $l_{a}$. The lateral distance from the attacker's position to the route is denoted in Equation~\ref{eq:la} below:
    \begin{equation} \label{eq:la}
        l_a = |\vec{p} - \vec{q}|
    \end{equation}

    \item Adversarial Vehicle's heading error $\Delta\theta_a$. We have the angular difference between the attacker's heading $\theta_{a}$ and the route direction in Equation~\ref{eq:theta-a}:
    \begin{equation} \label{eq:theta-a}
        \Delta\theta_a = \theta_a - \text{atan2}(y_{i+1} - y_{i}, \ x_{i+1} - x_{i})
    \end{equation}
\end{icompact}


\paragraph{Victim Vehicle Relative Status:} 

\sys monitors the relative position of the victim vehicle using a rear-facing camera mounted on the adversarial vehicle. Specifically, a real-time object detector identifies the victim vehicle in the camera frame with a bounding box $\mathbf{B} = {(u, v, w, h)}$, where ${u, v}$ is the coordination of the center of the box and $(w, h)$ are the width and height in pixels.  Then, \sys calculates the victim vehicle's status using the following variables: 

\begin{icompact}
\item Relative distance $d_v$ between the victim and the adversarial vehicles. \sys estimates the distance $d_v$ to the victim from the bounding box size, assuming a known reference vehicle width $w_{v}$ and $f$ the focal length of the camera in pixels, we have a definition of $d_v$ in Equation~\ref{eq:dv}:
\begin{equation} \label{eq:dv}
d_v = \frac{f \cdot w_v}{w}
\end{equation}

\item Heading error ($\Delta\theta_v$) of the victim relative to the adversary's centerline. 
Assuming the victim's heading error as $\Delta\theta_v$ and the center of the entire rear image as $u_c$, we have Equation~\ref{eq:thetav}:
\begin{equation} \label{eq:thetav}
    \Delta\theta_v = \arctan\left(\frac{u - u_c}{f}\right)
\end{equation}
\end{icompact}








\paragraph{Patch Selection:}


At each time step, \sys selects a patch from the patch bank $\mathbf{P}$ based on where the victim is currently relative to the adversarial route. This requires two inputs: the victim's lateral offset $l_v$ from the adversarial route (Equation~\ref{eq:lv}) and its heading error $\Delta\theta$ relative to the direction of the route (Equation~\ref{eq:theta}): 
\begin{equation}\label{eq:lv}
    l_v = l_a + d_v \cdot \sin(\Delta\theta_v)
\end{equation}
\begin{equation}\label{eq:theta}
    \Delta\theta = \Delta\theta_a + \Delta\theta_v
\end{equation}
where $l_a$ and $\Delta\theta_a$ are the offset and heading error of the adversarial vehicle (Equations~\ref{eq:la}--\ref{eq:theta-a}), and $d_v$ and $\Delta\theta_v$ are the relative distance and heading error of the victim (Equations~\ref{eq:dv}--\ref{eq:thetav}). 

Given $l_v$ and $\Delta\theta$, \sys then computes a desired steering correction in Equation~\ref{eq:phi}:
\begin{equation}\label{eq:phi}
    \Delta\phi = -k_l \cdot l_v - k_\theta \cdot \Delta\theta
\end{equation}
where $k_l$ and $k_\theta$ are pre-set coefficients. Intuitively, this correction points the victim back toward the adversarial route. Lastly, \sys selects a patch from $\mathbf{P}$ that optimized the steering direction to match the most closely $\Delta\phi$ for the display on the back of the adversarial vehicle. 

\paragraph{Adversarial Vehicle Maneuver:} 
\sys also controls the adversary vehicle using a front-facing camera to monitor the road in real time. The objective is to preserve outwardly normal driving behavior for both the adversary and the victim, so that external observers and onboard passengers are unlikely to perceive anomalous behavior even while the victim is under attack.
In particular, when \sys detects regulatory cues such as a red traffic light or a stop sign, the adversary vehicle brakes to a compliant stop and temporarily disables the rear display (i.e., removes the adversarial patch), thus pausing the attack. In the absence of active influence, the victim vehicle reverts to its nominal driving policy and follows traffic control as usual. After the roadway becomes clear, the adversary vehicle resumes movement and re-establishes an effective following configuration by adjusting its distance and relative heading with respect to the victim, using the victim’s estimated position and orientation (Equations.~\ref{eq:lv} and~\ref{eq:thetav}). Once the appropriate geometry is restored, the adversary vehicle re-enables the rear display and renders the patch selected by the patch-selection module to continue the attack.

\section{Implementation}
\label{subsec:implementation}

We implemented an open-source prototype of \sys in approximately 3,400 lines of Python code. This implementation consists of two main components that coordinate with each other to adversarially hijack the victim vehicle. We describe the implementation details below:

\paragraph{Patch Optimizer (The offline attack execution):} 

As described in Section~\ref{subsec:stage1}, \sys optimizes adversarial patches using the Adam optimizer with a learning rate of $2/255$ for training and $0.5/255$ for fine-tuning. Each patch is optimized for 1,000 iterations, with fine-tuning initialized from an existing patch. We use Adam because patch optimization is high-dimensional and highly non-convex, and the gradients can be noisy due to stochastic sampling of driving frames and contextual variations such as viewpoint and illumination. Adam's per-parameter adaptive learning rates and momentum terms help stabilize optimization under heterogeneous gradient scales, enabling faster convergence from random initialization while reducing sensitivity to manual step-size tuning.
Training data consists of street-view images and GPS signals collected via the various routes. \sys trains patches for five target steering directions, $\{-18, -6, 0, 6, 18\}$ degrees, where $0$ indicates going straight, negative values indicate steering left, and positive values indicate steering right. For each steering direction, six frames are sampled from the collected driving data. The patch optimizer is implemented in approximately 1,400 lines of Python using PyTorch 2.3~\cite{pytorch23}, OpenCV 4.9~\cite{opencv49}, and Pillow 10.3~\cite{pillow103}.

\paragraph{Interactive Adjustment Loop (The online attack execution):} 

As described in Section~\ref{subsec:stage2}, the interactive adjustment loop operates as a lightweight feedback controller during the online attack. \sys applies a proportional gain of $k_l = 0.3$ to compute a steering correction from the lateral offset and heading error of the victim vehicle, and then selects the closest matching patch from the patch bank. To avoid frequent patch changes caused by small fluctuations, \sys uses a switch threshold $4^\circ$; that is, the patch selector updates the displayed patch only when the desired steering angle changes by more than approximately $4^\circ$. The interactive adjustment loop is implemented in approximately 600 lines of Python code using NumPy 1.26~\cite{numpy,numpy126}.

\paragraph{Adversarial and Victim Autonomous Vehicles:}


The victim AV is controlled by an end-to-end autonomous driving agent equipped with a front-facing RGB camera ($1920 \times 960$ resolution, $110^{\circ}$ FOV), a Global Navigation Satellite System (GNSS) receiver, an Inertial Measurement Unit (IMU), and a speedometer. The agent processes camera frames and measurements from auxiliary sensors through a vision encoder and a multimodal fusion module. The fused representation is then passed to the decision model, which outputs steering, throttle, and brake commands at each time step.
The adversarial vehicle operates in autopilot mode and maintains a position approximately $12\,\mathrm{m}$ ahead of the victim vehicle in its forward direction. The autopilot agent continuously adjusts the adversarial vehicle's speed to match the victim's velocity while preserving the target following distance. The adversarial and victim vehicle components are implemented in approximately 1,500 lines of Python code using PyTorch 2.2~\cite{pytorch22}, Transformers 4.46~\cite{transformers446}, and NumPy 1.23~\cite{numpy123}.

\section{Evaluation}
\label{sec:eval}

In this section, we first present our experimental setup in Section~\ref{subsec: setup}, followed by the evaluation results for each research question in Sections~\ref{sec:eval:rq1}--\ref{sec:eval:rq5}.

\subsection{Experimental Setup} \label{subsec: setup}

\noindent{\bf Hardware:}
We conducted our digital experiments on a GPUs cluster equipped with NVIDIA A100~(80 ~ GB) GPUs, using PyTorch~2.3 with CUDA~12.5. The Stage~2 online experiments are run on a local workstation with a single NVIDIA RTX~3090 GPU, which maintains a real-time camera processing rate of approximately 20--30~Hz.


\vspace{0.05in}
\noindent{\bf Victim Autonomous Vehicles:}
We evaluate \sys against two driving agents of the victims. The first is SimLingo~\cite{simlingo}, a vision-language action driving model built on the InternVL2-1B backbone. The InternViT-based vision encoder extracts image features from front-camera views and a LoRA-adapted Qwen2 language backbone to fuse visual, navigation, ego-state, and prompt information for both language reasoning and action prediction.
The second is TCP~\cite{tcp}, a trajectory-conditioned planning model that uses a ResNet-34 encoder with 26.6\,M parameters and a GRU-based waypoint decoder. Both agents predict future trajectory waypoints, which are then converted into steering and throttle commands using PID controllers. We use the default hyperparameter settings for both models. All simulation experiments are conducted in CARLA~0.9.15.


\vspace{0.05in}
\noindent{\bf Autonomous Driving Datasets and Benchmarks:}  
Our evaluation is based on Bench2Drive~\cite{Jia2024Bench2Drive_NeurIPS}, a suite of scenario-driven closed-loop tasks designed to stress-test perception, reasoning, and planning under realistic driving conditions. We randomly sample 13 benign routes from the benchmark without modification, selecting routes in town maps that contain multiple intersections. For each benign route, we generate three adversarial routes from the same starting point. At each intersection, we choose alternative directions that differ from both the benign route and the direction from which the victim vehicle entered the intersection. An adversarial route terminates at the second intersection, whereas the remaining adversarial routes continue to the next intersection, where the same procedure is repeated to select additional alternative directions. As a result, each benign route is paired with three adversarial routes containing one to three intersections, and the adversarial turning direction at each intersection differs from the corresponding benign route.


\begin{figure}[t]
  \centering
  \begin{subfigure}{0.48\columnwidth}
    \centering
    \includegraphics[width=\linewidth]{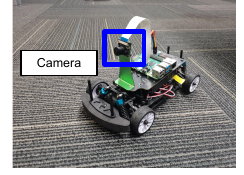}
    \caption{Victim Vehicle}
  \end{subfigure}
  \hfill
  \begin{subfigure}{0.48\columnwidth}
    \centering
    \includegraphics[width=\linewidth]{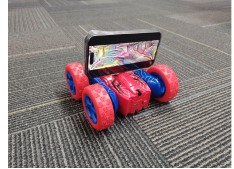}
    \caption{Attacker Vehicle}
  \end{subfigure}

  \caption{Physical attack setup. 
    (a) The victim car is equipped with a front-facing camera and remotely controlled by a TCP model running on a server. 
    (b) The attacker car is equipped with a screen displaying the adversarial patches.}
  \label{fig:physical_setup}
\end{figure}

\begin{figure}[t]
  \centering
  \begin{subfigure}{0.49\columnwidth}
    \centering
    \includegraphics[width=\linewidth]{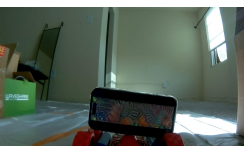}
    \caption{Victim front view in environment~1}
  \end{subfigure}
  \hfill
  \begin{subfigure}{0.49\columnwidth}
    \centering
    \includegraphics[width=\linewidth]{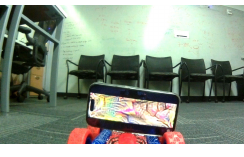}
    \caption{Victim front view in environment~2}
  \end{subfigure}

  \caption{
    Victim's camera views in two different physical environments}
  \label{fig:physical_frames}
\end{figure}

\subsubsection{Physical Environment}


We implemented our physical autonomous driving platform using a Donkey Car equipped with a Raspberry Pi and a front-facing PiCamera ($1920 \times 1080$)~\cite{donkeycar}. The victim vehicle is controlled by a TCP agent running on a separate server: each camera frame is streamed over WiFi to the server, which returns the corresponding steering command. The attacker vehicle drives ahead of the victim and displays adversarial patches on a phone screen mounted on its rear, as shown in Figure~\ref{fig:physical_setup}. We evaluated the attack in two indoor settings with different floor textures, lighting conditions, and track layouts, as shown in Figure~\ref{fig:physical_frames}.

\subsubsection{Evaluation Metrics}
We use the following metrics in our experiments.

\begin{icompact}

\item{\it Hijacking Success Rate (HSR):}
A hijack is considered successful if the victim vehicle reaches the adversary-designated area within a small radius $r$ (e.g., 6 meters) before a timeout $T$ (e.g., 10 minutes). HSR is the percentage of successful hijacks over all trials.

\item{\it Hijacking Compliance Rate (HCR):}
HCR measures how closely the victim follows the adversarial route. It is defined as the ratio between the maximum distance traveled by the victim along the adversarial route and the total length of the adversarial route.

\item{\it Steering Compliance Rate (SCR):}
SCR measures whether the victim's steering response follows the intended direction after the attacker switches to a new patch. We smooth the steering signal using a $w$-frame moving average ($w=10$, corresponding to 0.5\,s at 20\,Hz). For each patch switch, the response is considered \emph{compliant} if the victim's smoothed steering changes in the same direction intended by the new patch. SCR is the fraction of compliant patch switches.

\item{\it Traffic Rule Violations:}
This metric counts the number of traffic rule violations committed during a trial, including red-light and stop-sign violations. We report the total number of violations and separate them by violation type.

\item{\it Trajectory Curvature:}
Trajectory curvature measures how sharply the vehicle's path bends. It is defined as the rate of change of the trajectory's tangent angle with respect to arc length.

\item{\it Number of Hard Brakes:}
A hard brake is a sudden and strong deceleration event. Following common definitions~\cite{hardbrake}, we define a hard brake as a deceleration greater than $3\,\mathrm{m/s^2}$, approximately $6.7\,\mathrm{mph/s}$.

\item{\it Number of Steering Reversals:}
A steering reversal occurs when the steering input changes direction, such as switching from left steering to right steering or vice versa.

\end{icompact}

\subsubsection{Research Questions:}
In the remainder of the evaluation, we answer the following research questions:

\begin{icompact}
  \item {\bf RQ1: Attack Reliability:} Can \sys reliably hijack the victim vehicle to an adversary-designated destination over long driving horizons?

  \item {\bf RQ2: Attack Robustness:} How robust is \sys under environmental shifts, such as changes in weather and lighting conditions?

  \item {\bf RQ3: Attack Stealthiness:} Does the victim vehicle maintain normal-looking driving behavior during hijacking, similar to an unattacked autonomous vehicle?

  \item {\bf RQ4: Resilience against Defenses:} Can \sys remain effective against defenses designed to detect or mitigate adversarial patch attacks?

  \item {\bf RQ5: Attack Transferability:} Can adversarial patches trained on one driving model remain effective against another model with different weights or action heads?

  \item {\bf RQ6: Physical-World Attack:} Can \sys hijack a physical autonomous vehicle in real-world settings?
  
\end{icompact}


\subsection{RQ1: Overall Hijacking Reliability}
\label{sec:eval:rq1}

For RQ1, we evaluate the overall hijacking reliability of \sys in terms of HSR, HCR, and SCR across all adversarial routes in the simulation environment. 
Table~\ref{tab:rq1-main} summarizes the results. \sys successfully hijacks 34 of 39 routes for both SimLingo and TCP, achieving average HCRs of 91.4\% and 92.9\%, respectively. The SCR is 73.3\% for SimLingo and 78.8\% for TCP, indicating that the victim vehicle's steering mostly follows the attacker's intended directional changes during the hijack. We also observe that the HSR decreases as the number of intersections increases. This is because longer routes require the attacker to maintain control over more decision points, while the victim vehicle continuously attempts to recover and return to its originally planned route. 


The failure cases across both models are mostly attributed to environmental factors. Some failures are caused by software crashes in the CARLA simulation framework, likely triggered by edge-case interactions between the attacker vehicle and specific map geometries.  Other failures occur at signalized intersections where red-light timing and background vehicle behavior interfere with the attacker's ability to maintain positioning in front of the victim.  The remaining failures involve narrow road segments in certain CARLA towns where the map layout leaves insufficient room for the formation of two-vehicles.  Despite these environmental challenges, the overall HSR of 34/39 for both models demonstrates that \sys reliably hijacks the victim on the majority of adversarial routes.

\subsubsection{A Case Study}
We use SimLingo as the victim model in the following case study, as well as in RQ2 and RQ3, since both models exhibit similar attack success rates (Table~\ref{tab:rq1-main}).

\begin{figure}[t]
    \centering
    \includegraphics[width=\linewidth]{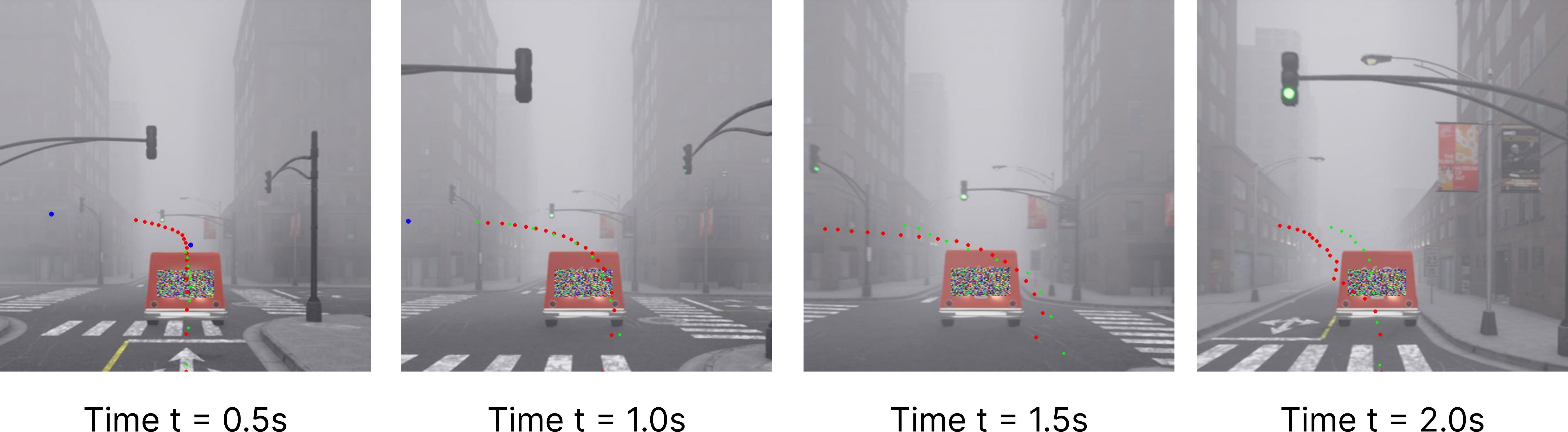}
    \caption{[RQ1] In this case study, the victim vehicle is supposed to turn left at the intersection, but is redirected to go straight by \sys. The red and green dots are the victim vehicle's predicted trajectories. The red dots are path waypoints depicting the victim's future position, and green dots are speed waypoints depicting the victim's future speed.}
    \label{fig:case_studies}
\end{figure}

Figure~\ref{fig:case_studies} shows a successful hijack on a one-intersection route. The benign route of the victim vehicle requires a turn left at the upcoming intersection, while the adversarial route instructs the victim to continue straight. The hijack scenario starts as follows:
At $t{=}0.5s$, the adversarial vehicle merges in front of the victim and displays an initial patch for straight-line following. As both vehicles approach the intersection ($t{=}1s$), the interactive adjustment loop detects a slight leftward drift in the victim's heading, which is consistent with its internal navigation that planned to turn left.  In response, the interactive loop switches to a patch optimized for a right-hand steering correction of approximately $6^{\circ}$ to counteract the victim's intent. At $t{=}1.5s$, the victim passes the intersection going straight and is fully committed to the adversarial route. The adversarial vehicle then returns to the neutral patch ($0^{\circ}$) to maintain lane centering. At $t=2s$, the victim arrives within the 6-meter acceptance radius of the adversary-designated destination and \sys completes the hijack. Throughout the process, the victim obeys all traffic signals (green light), maintains a comfortable speed profile, and does not make sharp steering changes, demonstrating that \sys is effective and stealthy.

\begin{table}[t]
  \centering
  \footnotesize
  \caption{[RQ1] Hijacking reliability results across two victim models. HSR: Hijacking Success Rate, HCR: Hijacking Compliance Rate, SCR: Steering Compliance Rate.}
  \label{tab:rq1-main}
  \setlength{\tabcolsep}{6.5pt}
  \renewcommand{\arraystretch}{1.05}

  \begin{tabular}
  {@{\extracolsep{\fill}} l ccc ccc}
    \toprule
    \multirow{2}{*}{\textbf{\# Int.}}
      & \multicolumn{3}{c}{\textbf{SimLingo}}
      & \multicolumn{3}{c}{\textbf{TCP}} \\
    \cmidrule(lr){2-4} \cmidrule(lr){5-7}
      & \textbf{HSR} $\uparrow$ & \textbf{HCR} $\uparrow$ & \textbf{SCR} $\uparrow$
      & \textbf{HSR} $\uparrow$ & \textbf{HCR} $\uparrow$ & \textbf{SCR} $\uparrow$ \\
    \midrule
    1       & 12/13 & 95.4$\pm$8.7  & 70.8
            & 13/13 & 100$\pm$0.0   & 83.8 \\
    2       & 12/13 & 93.7$\pm$11.9 & 73.1
            & 11/13 & 91.5$\pm$19.8 & 73.8 \\
    3       & 10/13 & 85.1$\pm$17.0 & 75.8
            & 10/13 & 87.3$\pm$12.6 & 76.2 \\
    \midrule
    Overall & 34/39 & 91.4$\pm$7.6  & 73.3
            & 34/39 & 92.9$\pm$5.0  & 77.9 \\
    \bottomrule
  \end{tabular}
\end{table}

\subsection{RQ2: Environmental Shifts}
\label{sec:eval:rq2}

\begin{figure}[!t]
  \centering
  \includegraphics[width=\linewidth]{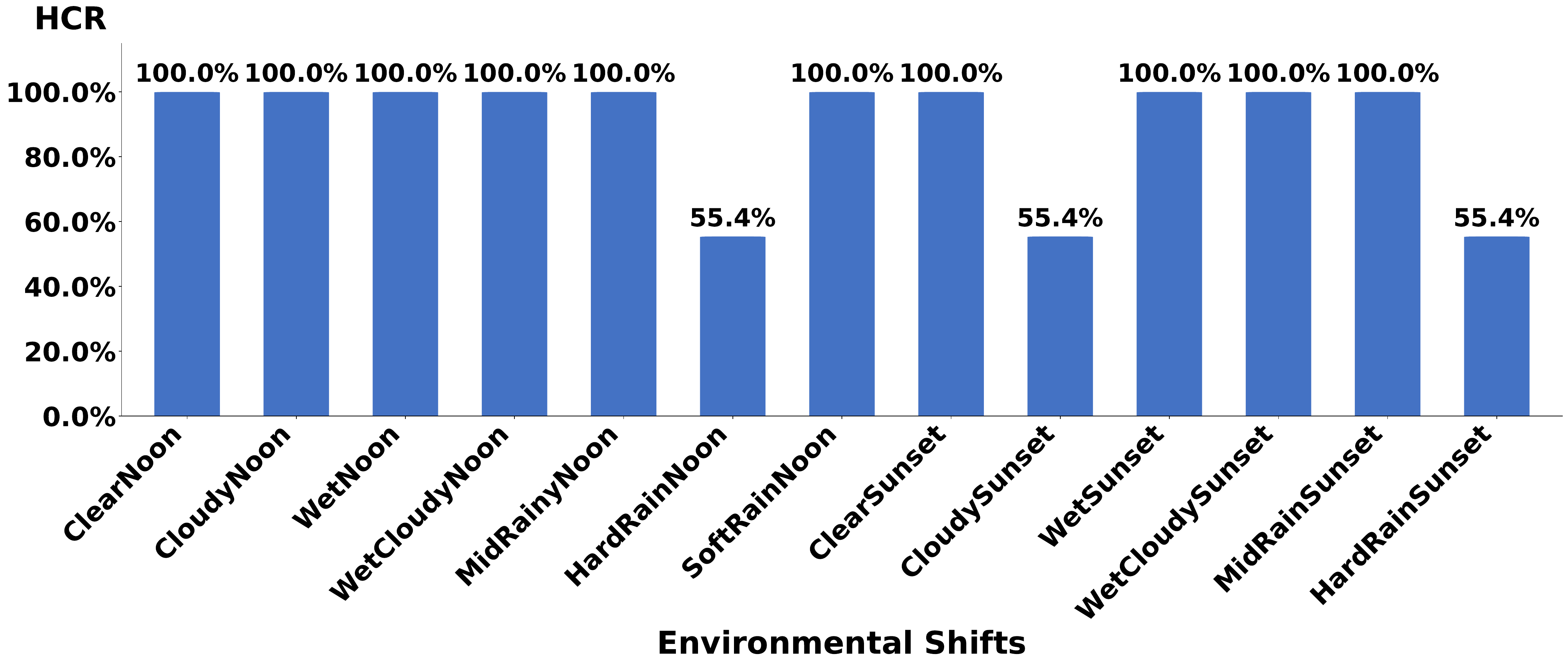} 
  \caption{[RQ2] HCR across different environmental shifts. 
  } \label{fig:rq2weather}
\end{figure}

For RQ2, we evaluate how environmental changes affect the hijacking compliance rate on a randomly selected adversarial route. We consider two major factors: (i) weather and lighting conditions, and (ii) traffic volume.
First, following the setting of previous work~\cite{song2023usenix}, we evaluated \sys under 14 environmental conditions that cover different combinations of weather and lighting. We vary four weather-related parameters: sun altitude angle $(S)$, cloudiness $(C)$, ground precipitation $(P_g)$, and airborne precipitation $(P_a)$. Figure~\ref{fig:rq2weather} reports the HCR under these conditions. \sys achieves a 100\% hijacking compliance rate in all but three scenarios.

\begin{figure}[!t]
    \centering
    \begin{subfigure}[b]{\linewidth}
        \centering
        \includegraphics[width=\textwidth]{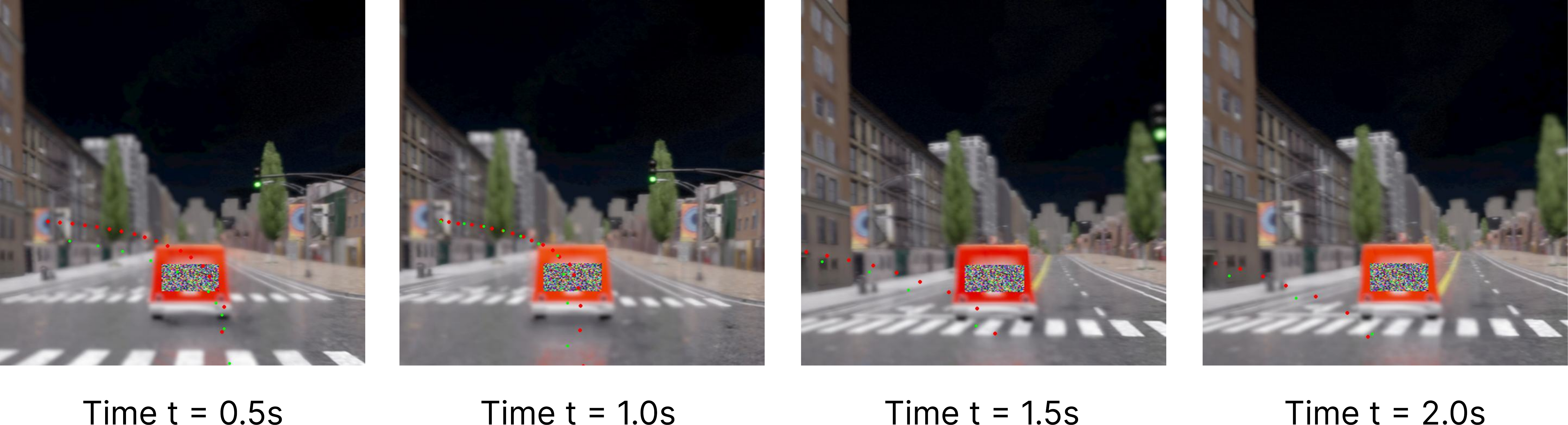}
        \caption{Hard rain noon}
    \end{subfigure}
    
    \vspace{0.2cm}
    
    \begin{subfigure}[b]{\linewidth}
        \centering
        \includegraphics[width=\textwidth]{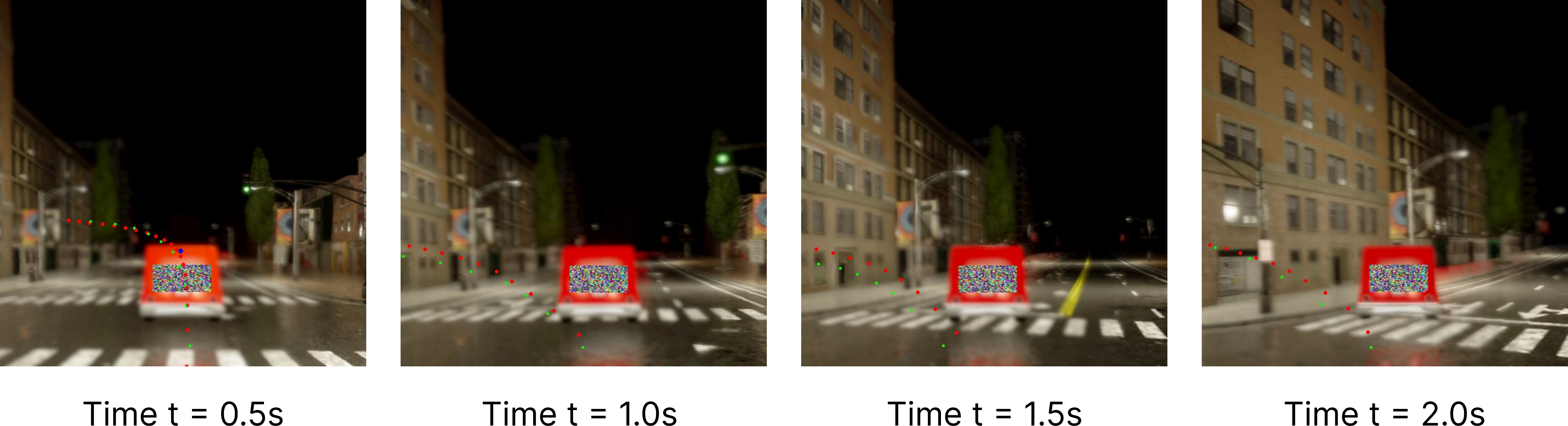}
        \caption{Hard rain sunset}
    \end{subfigure}
    
    \vspace{0.2cm}
    
    \begin{subfigure}[b]{\linewidth}
        \centering
        \includegraphics[width=\textwidth]{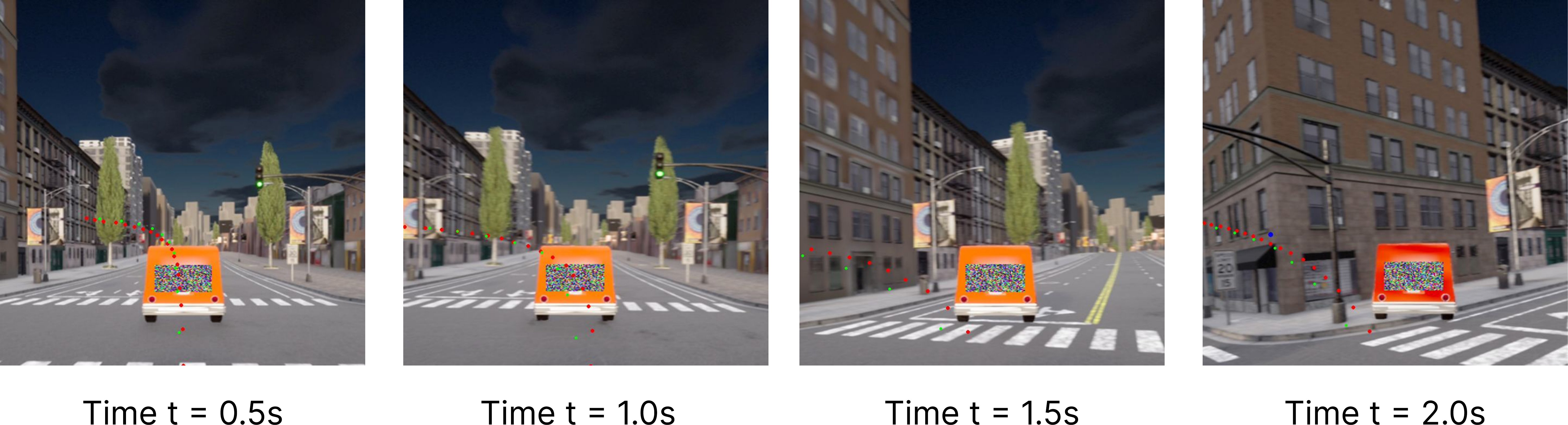}
        \caption{Cloudy sunset}
    \end{subfigure}
    \caption{[RQ2] An illustration of failed hijacks under different 
    environmental shifts.}
    \label{fig:rq2failure}
\end{figure}

Figure~\ref{fig:rq2failure} illustrates these three failure scenarios. We identify two main failure modes:
\begin{icompact}
\item \textit{Perception blur caused by heavy rain.} 
Figures~\ref{fig:rq2failure}(a) and (b) show failures under heavy rain at noon and sunset, respectively. In both cases, high ground precipitation $P_g$ and airborne precipitation $P_a$ significantly degrade visual quality. Water accumulation and rain streaks blur the victim vehicle's front-camera view, while precipitation between the two vehicles introduces motion blur that obscures patch details. As a result, the victim camera cannot clearly capture the adversarial patch, reducing its effectiveness.
\item \textit{Color shift caused by lighting.} 
Figure~\ref{fig:rq2failure}(c) shows a failure under cloudy sunset conditions. Unlike the heavy-rain cases, the victim vehicle has a clear view of the adversarial vehicle. However, the low sun altitude angle $S$ creates strong directional reflection from the rear of the adversarial vehicle. As shown at $t=0.5$ and $t=1$, when the adversarial vehicle moves through the intersection, its orientation relative to the sun changes and produces a specular reflection. From the victim's perspective, this reflection causes a temporary color shift, especially over the adversarial patch. This dynamic color shift changes the feature distribution perceived by the victim's vision encoder, weakening the adversarial context that the patch was optimized to induce.
\end{icompact}

Second, we evaluate the robustness of \sys under different traffic volumes. Specifically, we tested three traffic settings: heavy traffic with 4 vehicles per lane and 12 vehicles per junction, medium traffic with 2 vehicles per lane and 6 vehicles per junction, and light traffic with 1 vehicle per lane and 2 vehicles per junction. Table~\ref{tab:rq2traffic} reports the HCR under these traffic conditions. The results show that \sys can still successfully hijack the victim vehicle even in heavy traffic. This is because the offline Min--Max optimization stage accounts for variations in traffic density, allowing the selected patches to remain effective despite changes in the number of surrounding vehicles.


\begin{table}[!t]
  \centering \footnotesize
  \setlength{\tabcolsep}{6pt}
  \caption{[RQ2] HCR across different traffic volumes. Criteria columns 
  show the number of vehicles per lane (\#Veh./Lane), per junction 
  (\#Veh./Junction), and the spacing between vehicles. \sys achieves 100\% HCR for all scenarios.}
  \label{tab:rq2traffic}
  \small
  \begin{tabular}{@{} l ccc c @{}}
    \toprule
    \multirow{2}{*}{\textbf{Traffic}} & \multicolumn{3}{c}{\textbf{Criteria}} & \multirow{2}{*}{\textbf{HCR (\%)} $\uparrow$} \\
    \cmidrule(lr){2-4}
    & \textbf{\#Veh./Lane} & \textbf{\#Veh./Junction} & \textbf{Spacing} & \\
    \midrule
    Light  & 2 & 2  & 25m & 100 \\
    Medium & 4 & 6  & 15m & 100 \\
    Heavy  & 8 & 12 & 10m & 100 \\
    \bottomrule
  \end{tabular}
\end{table}

\subsection{RQ3: Stealthiness}
\label{sec:eval:rq3}

For RQ3, we evaluate the stealthiness of the hijacking attack. Our evaluation compares the vehicle behavior of the victim under \sys with benign driving behavior, where the victim is intentionally routed along the same adversarial route. The closer the hijacked behavior is to this benign baseline, the stealthier the attack appears since an observer may interpret the victim's motion as a normal route following rather than adversarial manipulation. Specifically, we evaluate stealthiness using two categories of metrics: (i) trajectory smoothness and unsafe driving behaviors along the route, and (ii) the number of traffic-rule violations.

\begin{table}[!t]
\centering \footnotesize 
\setlength{\tabcolsep}{7pt}
\caption{[RQ3] A comparison of trajectory smoothness and unsafe driving behaviors between benign driving and adversarial hijacking under the influence of \sys.}
\label{tab:trajectory}
\begin{tabular}{lcccc}
\toprule
\multirow{2}{*}{\textbf{Metric}} & \multicolumn{2}{c}{\textbf{Benign Driving}} & \multicolumn{2}{c}{\textbf{Adversarial Hijacking}} \\
\cmidrule(lr){2-3} \cmidrule(lr){4-5}
& Average & Range & Average & Range \\
\midrule
Curvature $\downarrow$        & $3.47$ & $[0.20, 36.79]$ & $2.89$ & $[0.51, 9.17]$ \\
Hard Brakes $\downarrow$      & $97.46$ & $[19, 300]$ & $43.62$ & $[19, 89]$ \\
Steer Reversals $\downarrow$  & $0.20$ & $[0.01, 0.64]$ & $0.19$ & $[0.06, 0.32]$ \\
\bottomrule
\end{tabular}
\end{table}

Table~\ref{tab:trajectory} reports the average values and ranges for trajectory curvature, hard braking events, and steering reversals. Overall, adversarial hijacking runs exhibit values that are comparable to, and in several cases lower than, those observed during benign driving. This suggests that \sys does not introduce obvious instability or abnormal maneuvering. Looking at each metric individually, the curvature of the trajectory remains low because \sys guides the victim along adversarial routes that are generally smooth. The number of hard brakes is also small because the adversarial vehicle drives in front of the victim and attempts to maintain a stable following distance, reducing the need for sudden deceleration. Similarly, the number of steering reversals remains low because adversarial routes and the \sys control strategy are designed to avoid abrupt changes in steering direction, which are more difficult to reliably manipulate.

\begin{table}
\centering \footnotesize 
\setlength{\tabcolsep}{12pt}
\caption{[RQ3] A comparison of traffic rule violations between benign driving and adversarial hijacking under the influence of \sys.} \label{tab:trafficrules}
\begin{tabular}{lcc}
\toprule
{\bf Traffic Rule} & {\bf Benign Driving} & {\bf Adversarial Hijacking} \\
\midrule
Red Light $\downarrow$  & {0/7} & {0/7} \\
Stop Sign $\downarrow$  & {0/5} & {2/5} \\
\bottomrule
\end{tabular}
\end{table}


Moreover, Table~\ref{tab:trafficrules} compares traffic-rules violations during benign driving and adversarial hijacking on the same routes. Both settings result in zero red-light violations. However, adversarial hijacking leads to two stop-sign violations, whereas benign driving does not. This difference is due to the greater difficulty in maintaining stop-sign compliance when two vehicles follow each other. At a red light, both adversarial and victim vehicles stop and proceed once the light turns green. At a stop sign, both vehicles initially stop, but after the adversarial vehicle moves forward, the victim must perform an additional stop before proceeding to fully satisfy the rule. This behavior requires fine-grained control over the adversarial patches displayed. In both observed violations, the victim stopped once but did not complete the second stop. Although this is a violation of traffic-rules, it resembles a roll stop~\cite{rollingstop}, which is also commonly observed in human driving.


\begin{table}[!t]
  \centering
  \footnotesize
  \setlength{\tabcolsep}{20pt}
  \caption{[RQ4] Attack performance with and without LGS defense on TCP (L1 routes). 
  \sys is still mostly effective when LGS is deployed as a defense. } 
  \label{tab:defense}
  \begin{tabular}{@{} lccc @{}}
    \toprule
    \textbf{Setting} & \textbf{HSR} $\uparrow$ & \textbf{HCR (\%)} $\uparrow$ & \textbf{SCR (\%)} $\uparrow$ \\
    \midrule
    No defense  & 13/14 & 96.7$\pm$6.2  & 83.8 \\
    With LGS    & 9/14  & 92.1$\pm$7.1  & 81.9 \\
    \bottomrule
  \end{tabular}
\end{table}

\subsection{RQ4: Resilience against Defenses}
\label{sec:eval:rq4}

For RQ4, we evaluate whether existing adversarial patch defenses can mitigate our hijacking attack \sys. We selected Local Gradient Smoothing (LGS)~\cite{lgs} as our target defense for two reasons. First, it is the best fit specifically for this type of threat model: it identifies image regions with abnormally high gradient magnitudes, a common characteristic of adversarial patches, and suppresses them through local smoothing. Second, LGS is practical for real-time autonomous driving because it introduces only approximately ${\sim}10$\,ms of latency per frame, which remains within the timing budget of a 20\,Hz driving pipeline. In contrast, several other adversarial patch defenses introduce substantially higher latency~\cite{SAC, diffpad, diffpure}. For example, diffusion-based defenses, such as DiffPure~\cite{diffpure} require approximately ${\sim}200$\, milliseconds per frame, making them unsuitable for real-time autonomous driving.

Nevertheless, we integrated LGS into a TCP agent as an image preprocessing step. Before each camera frame is passed to the driving model, LGS scans the image for high-gradient regions and applies Gaussian smoothing to reduce their effect. We use a window size of $3$, a threshold percentile of $98$, a blur sigma of $0.2$, and a kernel size of $3$. We evaluate this defense on all 14 one-intersection routes using the same attack configuration as in RQ1.

In Table~\ref{tab:defense}, we compare attack performance with and without LGS defense.  Even when LGS is enabled, \sys maintains high attack compliance, achieving an HCR of 92.1\% and an SCR of 81.9\%, which are close to the undefended results of 96.7\% HCR and 83.8\% SCR. However, the HSR decreases from 92.9\% to 64.3\%. A closer analysis shows that one of the five failed routes also fails in the undefended setting, corresponding to the same challenging route discussed in RQ1. For the remaining four failures, the victim completes 98--99\% of the adversarial route before drifting slightly away from the attack path near the endpoint. This indicates that LGS only introduces minor degradation near the end of long routes, rather than effectively preventing the hijacking attack.

Thus, LGS introduces only 10.7\,ms of additional latency per frame on average. Our gradient analysis further shows that even after LGS preprocessing, the patch region retains a gradient magnitude $10.3\times$ higher than the surrounding background, indicating that the defense does not fully suppress the adversarial signal. This is because \sys's Min--Max Optimization Procedure, described in Section~\ref{subsec:stage1}, explicitly trains patches to remain effective under worst-case contextual perturbations, including variations in the background around the patch. Consequently, the optimized patches encode adversarial features that can survive the local gradient smoothing process done by LGS.


\subsection{RQ5: Blackbox Attack with Transferability}
\label{sec:eval:rq5}


In this research question, we evaluate the transferability of attacks under a blackbox setting as described in Section~\ref{subsec:threat} and Table~\ref{tab:attackercapability}. Specifically, we assume that the model weights and architecture deployed by a victim car may be different from those obtained by the adversary due to version or software updates to the vehicle's firmware. 



Our experiment setup is as follows.  We obtained two variants of the Trajectory-guided Control Prediction models: a Bench2DriveZoo (B2D) model~\cite{B2D} and an OpenDrive model~\cite{OpenDrive}.  These two models differ in architecture: While both use a ResNet-34 visual backbone, the OpenDrive model includes a GRU with different hidden dimensions and modified action heads. Then, we evaluate the patches obtained from the B2D model against a victim car using two different OpenDrive models, one original and the other that is fine-tuned based on the 4GB Bench2Drive Base Dataset~\cite{Jia2024Bench2Drive_NeurIPS} for 15 epochs.  
Our evaluation results are shown in Table~\ref{tab:transfer}. The patch trained based on the B2D model has 100\% HCR, and the SCR is also on par with the result when the adversary knows the model weights. This experiment shows that our patches have transferability under a blackbox setting. 
 


\begin{table}[!t]
  \centering
  \small
  \setlength{\tabcolsep}{2.5pt}
  \caption{[RQ5] Attack transferability results on four adversarial routes.  Patches are trained on the Bench2DriveZoom (B2D) models but applied to two different OpenDrive models.}
  \label{tab:transfer}
  \begin{tabular}{@{} l cc cc cc cc @{}}
    \toprule
    & \multicolumn{2}{c}{\textbf{Route 013}} & \multicolumn{2}{c}{\textbf{Route 020}} & \multicolumn{2}{c}{\textbf{Route 058}} & \multicolumn{2}{c}{\textbf{Route 098}} \\
    \cmidrule(lr){2-3} \cmidrule(lr){4-5} \cmidrule(lr){6-7} \cmidrule(lr){8-9}
    \textbf{Target} & HCR & SCR & HCR & SCR & HCR & SCR & HCR & SCR \\
    \midrule
    B2D $\to$ OpenDrive-1         & 100 & 100 & 100 & 84.72 & 100 & 89.19 & 100 & 73.22 \\
    B2D $\to$ OpenDrive-2    & 100 & 72.3 & 89.19 & 75.90 & 100 & 70.83 & 79.00 & 73.22 \\
    \bottomrule
  \end{tabular}
\end{table}

\subsection{RQ6: Physical-World Attack}
\label{sec:eval:rq6}




\vspace{0.04in}
\noindent{\bf Experiment Design:}
Our physical evaluation adopts a multi-phase experiment to evaluate whether \sys patches can reliably control the directional behavior of the victim vehicle. During each run, the victim vehicle is programmed to go straight always and \sys executes a predefined multi-phase patch sequence, switching between directional patches such as left, straight, and right to evaluate closed-loop directional control. There are two different environments with different lighting conditions.  In environment~1, we use a three-phase plan: left$\to$straight$\to$right. In environment~2, we use a two-phase plan: right$\to$straight.

\begin{figure}[!t]
  \centering
  \begin{subfigure}{\columnwidth}
    \centering
    \includegraphics[width=\columnwidth]{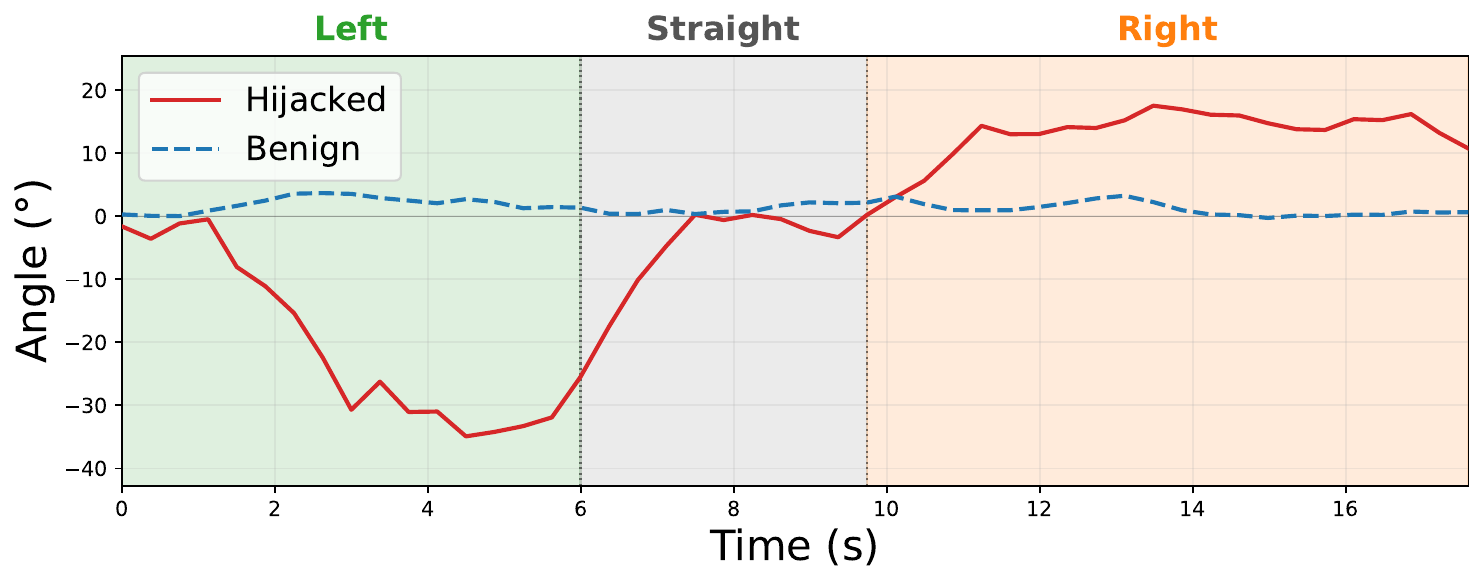}
    \caption{Steering angles using \sys in environment~1}
  \end{subfigure}

  \vspace{0.5em}

  \begin{subfigure}{\columnwidth}
    \centering
    \includegraphics[width=\columnwidth]{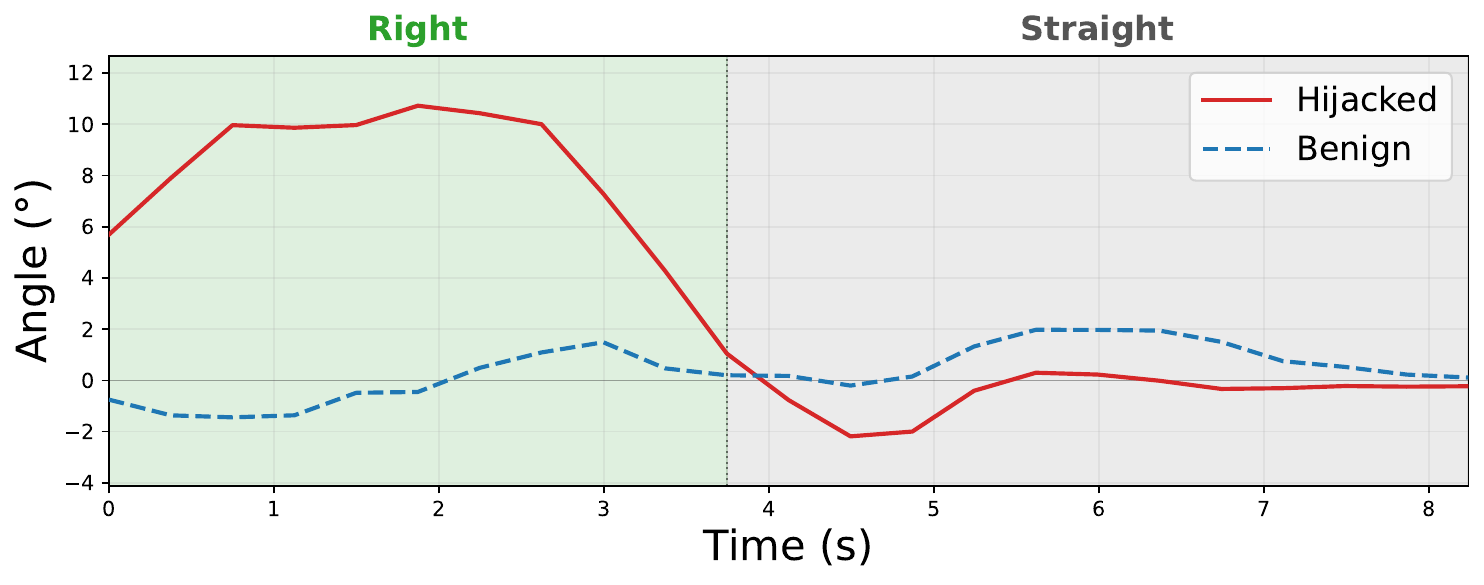}
    \caption{Steering angles using \sys in environment~2}
  \end{subfigure}

  \caption{
    [RQ6] Steering angle over time: benign driving (blue dashed) vs.\ hijacked driving (red solid).
    Negative values indicate left turning and positive right turning.
    Background colors indicate the active patch phase.
    The adversarial patch induces a clear, sustained steering deviation in each phase.}
  \label{fig:steering_timeseries}
\end{figure}

\vspace{0.04in}
\noindent{\bf Results:}
Figure~\ref{fig:steering_timeseries} compares the steering angles in benign driving (blue dashed line) and adversarial driving (red) in both environments. Under benign driving, the steering angle remains close to zero, with average values of $+1.4^\circ$ and $+0.3^\circ$ in environment~1 and environment~2, respectively. In contrast, under adversarial driving, the steering response changes consistently with the active patch. In environment~1 (Figure~\ref{fig:steering_timeseries}a), the left patch induces an average steering angle of $-20.5^\circ$, with a peak deviation of $-50^\circ$; the straight patch returns the steering angle close to zero; and the right patch shifts the steering angle to an average of $+13.1^\circ$. In the environment~2 (Figure~\ref{fig:steering_timeseries}b), the right patch produces a sustained average steering deviation of $+10.0^\circ$, while the straight patch brings the steering angle back to $+0.2^\circ$, closely matching the benign baseline. Table~\ref{tab:physical} summarizes the mean steering angle for each phase.

\sys achieves an HCR of 100\% in both environments, indicating that the victim vehicle completes the full adversarial protocol in each setting. The SCR is 83.3\% in environment~1 and 100\% in environment~2, as reported in Table~\ref{tab:physical}, showing that the victim's steering response aligns with the attacker's intended direction most of the time. These results demonstrate that \sys remains effective on a physical platform despite real-world deployment factors, including lighting variation, camera noise, and the display of the patch on a phone screen rather than as a printed pattern. This robustness is supported by the physical-world augmentations used during patch optimization, including brightness and contrast variation, scale jitter, and Gaussian blur.

\begin{table}[t]
  \centering
  \caption{[RQ6] Physical-world attack results on a Donkey Car platform with a TCP 
    driving agent. Steering angles are reported in physical degrees
    ($\pm60^\circ$ full range). SCR is computed as described in Section~\ref{subsec: setup} with $w{=}3$ frames.}
  \label{tab:physical}
  \footnotesize
  \setlength{\tabcolsep}{8.5pt} 
  \renewcommand{\arraystretch}{1.1}
  \begin{tabular}
  {@{\extracolsep{\fill}} llcrr}
    \toprule
    \textbf{Env.} & \textbf{Phase} & \textbf{Patch} & \textbf{Mean Angle (°)} & \textbf{SCR (\%)} \\
    \midrule

    \multirow{5}{*}{Env.1}
      & Benign (no patch)   & ---      & $+1.4$  & --- \\
      \cmidrule(l){2-5}
      & Phase 1 (0--5.6\,s)  & Left     & $-20.5$ & 75.0 \\
      & Phase 2 (6.0--9.4\,s) & Straight & $-5.3$  & 80.0 \\
      & Phase 3 (9.7--17.6\,s) & Right    & $+13.1$ & 90.9 \\
      \cmidrule(l){2-5}
      & \textbf{Overall}     & ---      & ---     & \textbf{83.3} \\
    \midrule

    \multirow{4}{*}{Env.2}
      & Benign (no patch)   & ---      & $+0.3$  & --- \\
      \cmidrule(l){2-5}
      & Phase 1 (0--3.4\,s)    & Right    & $+10.0$ & 100.0 \\
      & Phase 2 (3.7--12.0\,s) & Straight & $+0.2$  & 100.0 \\
      \cmidrule(l){2-5}
      & \textbf{Overall}     & ---      & ---     & \textbf{100.0} \\
    \bottomrule
  \end{tabular}
\end{table}

\section{Related Work}
\label{sec:relatedwork}

\noindent{\bf Adversarial attacks on vision-based AV perception.}
Previous work has shown that camera-based autonomous driving pipelines can be manipulated through physically realizable adversarial inputs. Sato et al.~\cite{sato2021usenix} showed that benign road-surface perturbations can bias lane-centering systems, causing sustained lateral control errors and physical-world safety violations. Other attacks target high-impact perception primitives such as traffic-light recognition, depth estimation, and obstacle understanding. For example, Rolling Colors~\cite{yan2022rollingcolors} demonstrated laser-based attacks against traffic-light perception, while DoubleStar~\cite{zhou2022doublestar} studied long-range manipulation of monocular and stereo depth estimation for obstacle avoidance. Attacks on traffic-signal recognition further show that physical perturbations can remain effective across viewpoints and environmental variations~\cite{jia2022foolingeyes}. 

Recent studies also show that adding more sensors does not fully remove the attack surface. Coordinated physical perturbations can induce consistent errors across camera and LiDAR streams~\cite{cao2021invisible}, and black-box LiDAR spoofing can be shaped to remain compatible with camera observations to bypass fusion-level checks~\cite{Hallyburton_USENIXSec2022}. Other attacks exploit the camera pipeline itself, runtime assumptions, or perspective cues, as shown by GlitchHiker~\cite{jiang2023glitchhiker}, $\pi$-Jack~\cite{zheng2024pijack}, and AEmorpher~\cite{zhu2024aemorpher}. Muller et al.~\cite{muller2025latency} further show that projector-based perturbations can increase detector latency and create a denial-of-service channel. These works motivate evaluating attacks under realistic physical constraints, temporal effects, and closed-loop driving outcomes rather than only frame-level perception accuracy.

\noindent{\bf Non-vision attacks on autonomous vehicles.}
Autonomous vehicles are also vulnerable to non-camera attacks that exploit in-vehicle networks, planning logic, and active sensors. At the network layer, CANflict~\cite{tron2022canflict} shows how attackers can leverage peripheral conflicts to mount data-link layer attacks on vehicle networks. Related defense and diagnosis systems, such as ZBCAN~\cite{serag2023zbcan} and RIDAS~\cite{shin2023ridas}, highlight the need for practical protection under legacy CAN constraints. At the planning layer, Wan et al.~\cite{wan2022tooafraid} identify semantic denial-of-service vulnerabilities in autonomous driving planners, where seemingly benign physical-world conditions can force conservative planners into mission-degrading behavior, such as persistent stopping. Other work shows that an adversarial vehicle can trigger unsafe outcomes through plausible interactive motion alone~\cite{song2023usenix}. LiDAR-focused attacks further demonstrate that adversaries can spoof or remove physical objects from the perceived scene~\cite{Cao_USENIXSec2023}, with recent work improving realism under long-distance and high-speed conditions~\cite{sato2025lidarrealism}. Together, these studies show that AV security must consider vehicular networks, planning and control logic, and the physical properties of non-camera sensors.

\noindent{\bf Adversarial attacks on drones.}
Adversarial attacks on Unmanned Autonomous Vehicles (UAVs) show that learning-based autonomy is vulnerable beyond ground vehicles. In aerial systems, perception errors can directly affect navigation, tracking, and collision avoidance. Closely related to our setting, Hanfeld et al.~\cite{10416782} demonstrated that attacker-controlled multirotors can carry optimized flying adversarial patches and place them inside a victim drone's field of view to manipulate its learned perception and control behavior. This mobile-patch setting is similar in spirit to \sys because the adversarial object is not fixed in the environment; it is carried by another agent that can maintain visibility during the attack. Other UAV-focused attacks study physical adversarial patches for object hiding, yaw manipulation, and obstacle avoidance~\cite{liu2024rpau}. These works emphasize the importance of the motion, viewpoint, distance, and physical deployment constraints. In contrast, \sys targets long-term route-level hijacking of vision-based AVs, where an adversarial vehicle must continuously influence the victim over multiple closed-loop perception--planning--control cycles while preserving realistic driving behavior.

\noindent{\bf Adversarial attacks on learning models.}
Autonomous driving systems also inherit broader adversarial ML threats, including backdoors and privacy leakage. Backdoor attacks are especially concerning because a model can behave normally on benign inputs while producing attacker-chosen behavior when a trigger appears. Previous work has studied backdoors in graph learning~\cite{xi2021graphbackdoor}, blind-label settings~\cite{bagdasaryan2021blindbackdoors}, explanation-guided poisoning~\cite{severi2021explanationguided}, and triggers of linguistic-style in NLP~\cite{pan2022lism}. Recent defenses target internal representations and poisoned data subsets, such as BEATRIX~\cite{ma2023beatrix} and ASSET~\cite{pan2023assetrobustbackdoordata}. As LLM and emerging architectures are integrated into safety-critical systems, backdoor risks also extend to code-generation models~\cite{yan2024codebreaker}, logic-based LLM backdoor detection~\cite{popovic2025debackdoor}, spiking neural networks~\cite{abad2024sneakyspikes}, and hardware fault attacks such as Rowhammer-based Trojan injection~\cite{li2025oneflip}. Privacy attacks, including enhanced membership inference~\cite{ye2022enhancedmia}, further show that deployed models can leak information about their training data. For autonomous driving, these threats imply that robustness must cover the full ML life cycle, from data collection and fine-tuning to deployment and updates.

\section{Conclusion}
This paper investigates a new class of physical-world attacks against vision-based autonomous vehicles: \emph{long-horizon route hijacking}. Unlike prior adversarial patch attacks that primarily trigger short-term failures, such as collisions or visible traffic-rule violations, route hijacking targets \emph{route integrity}---the ability of an AV to reach its intended destination by following the intended sequence of navigation decisions. To realize this threat, \sys treats visual patches as persistent \emph{steering primitives} rather than one-shot perturbations. It combines an offline optimization stage that constructs a patch bank robust to viewpoint and contextual variation with an online interactive adjustment loop that observes the victim vehicle's motion and dynamically switches among patches to sustain long-horizon influence under continual replanning. 
 Our results in both simulation and real-world environments show that physically realizable visual perturbations can be composed to achieve targeted route-level manipulation over extended driving horizons. We hope that our findings will motivate future defenses that reason over long-horizon consistency, including multi-sensor corroboration across vision, map priors, GNSS, and inertial cues.

\bibliographystyle{plain}
\bibliography{sample}

\end{document}